\documentclass[twocolumn,
aps,nofootinbib,showpacs,showkeys,preprint
tightenlines
] {revtex4}

\usepackage{epsf,epsfig,subfigure,axodraw,graphicx,amsmath,amssymb}
\usepackage{color}
\usepackage{float}
\newcommand{\dis}[1]{\begin{equation}\begin{split}#1\end{split}\end{equation}}

\newcommand{\tev}{\,\textrm{TeV}}
\newcommand{\gev}{\,\textrm{GeV}}

\newcommand{\NPQ}{$N_{\textrm{PQ}}$MSSM}
\newcommand{\MG}{{$M_{\rm GUT}$}}
\newcommand{\ie}{{\it i.e.}\ }
\newcommand{\etal}{{\it et al.}}

\begin{document}


\title{Singlet superfield extension of the minimal supersymmetric standard
model with Peccei-Quinn symmetry and a light pseudoscalar Higgs boson
at the LHC}

\author{Jihn E. Kim\email{jihnekim@gist.ac.kr}$^{1,3}$, Hans Peter Nilles$^2$, Min-Seok Seo$^3$
}
\address{
$^1$GIST College, Gwangju Institute of Science and Technology, Gwangju 500-712, Korea,
}
\affiliation{$^2$Bethe Center for Theoretical Physics and
Physikalisches Institut der Universit\"at Bonn, Nussallee 12, 53115 Bonn, Germany, and\\
 $^3$Department of Physics and Astronomy and Center for Theoretical
 Physics, Seoul National University, Seoul 151-747, Korea}
\begin{abstract}
Motivated by the $\mu$-problem and the axion solution to the strong
CP-problem, we extend the MSSM with one more chiral singlet field $X_{\rm ew}$.
The underlying PQ-symmetry allows only one more renormalizable term $X_{\rm ew} H_u H_d$ in
the superpotential. The  spectrum of the Higgs system includes a light
pseudoscalar $a_X$ (in addition to the standard CP-even Higgs boson),
predominantly decaying to two photons: $a_X \to \gamma \gamma$. Both
Higgs bosons might be in the range accessible to current LHC
experiments.
\end{abstract}


\maketitle

\section{Introduction}\label{sec:Introduction}

The Large Hadron Collider [LHC] at CERN in Geneva, Switzerland is the world's highest energy accelerator designed to study the physics relevant for the origin
of mass. According to the theory called the Standard Model [SM] which describes, with few exceptions, most of the observable phenomena in the universe, the Higgs field is responsible for the masses of the $W$ and $Z$ bosons, as well as that of quarks and leptons. Experiments, published so far, now only allow for the Higgs mass to be in the range between 115 and 127 GeV \cite{Atlas11,CMS11}.  The two humongous multi-purpose
detectors, CMS and ATLAS, at CERN have recently presented hints towards the discovery of the elusive Higgs boson \cite{Earler12}.  Details concerning its exact mass and branching ratios will tell us whether we are dealing with the Higgs
boson of the SM, or whether physics beyond the SM is required. At the moment all these possibilities are still open.

Supersymmetry is a mild extension of the SM. In its simplest form, the Minimal Supersymmetric Standard Model [MSSM], it favors a rather light CP-even
Higgs boson with mass below 130 GeV \cite{HiggsLoop} and properties very similar
to those of the SM Higgs boson. There are however two serious problems of the
MSSM.   These are known as the $\mu$ and strong CP problems.  The $\mu$ term ($\mu H_u H_d$,  where $H_u$ and $H_d$ denote the Higgs doublet superfields of the MSSM)
gives mass to the Higgs doublets of order $\mu$.  The problem is that this term
breaks no low energy symmetry and is thus naturally of order the $Planck$ scale, but it needs to be at the weak scale.   The strong CP problem is a problem of QCD.  One popular solution requires the existence of a light pseudo-scalar particle, known as the axion.

In order to address these problems simple extensions of the MSSM have been proposed.
The simplest (singlet) extension of the MSSM is the NMSSM  \cite{NSW}, motivated by questions of electroweak symmetry breakdown, the $\mu$-problem
and an increase of the upper limit on mass of the lightest
Higgs boson \cite{EllwangRev}. Properties of the Higgs system might
change drastically and could be checked by LHC experiments \cite{Ellwanger11}.
A relation between the $\mu$ problem and the
(invisible) axion solution \cite{InvAxions} to the strong CP-problem
have been noticed in a particular singlet extension \cite{KimNilles} of the MSSM.

In this letter we shall discuss a simple generalization \cite{Kim84}
of this scheme (which we denote by \NPQ) with
additional light supermultiplets, one of which ($X_{\rm ew}$) is protected
by the original Peccei-Quinn (PQ) symmetry \cite{PQ77}.
The PQ symmetry forbids the renormalizable terms, $H_uH_d, X_{\rm ew}, X_{\rm ew}^2$, and $X_{\rm ew}^3$. The coefficients of these can be generated by breaking the PQ symmetry at a high energy scale. In particular, we note that the coefficients of $X_{\rm ew}$ and $X_{\rm ew}^2$ are not the PQ symmetry breaking scale, in contrast to most PQ symmetry models where they are generically of order the PQ symmetry breaking scale. The $\mu$ term can be generated by a nonrenormalizable term, which is the one used in the MSSM. With the singlet $X_{\rm ew}$ surviving down to the TeV scale, there is an important renormalizable term $X_{\rm ew} H_u H_d$ which distinguishes the phenomenology of \NPQ~ from that of the MSSM.

The main result of this letter is the observation that such a model predicts the existence of a light pseudoscalar (CP-odd) Higgs
boson that could be within reach of the current LHC experiments. In the limit $\mu\sim 0$, we find that there exists a chiral symmetry which is nontrivial only for $H_u, H_d$ and $X_{\rm ew}$. It may be called `Higgsino symmetry'. Because of its pseudoscalar nature, such an (axion-like) particle $a_X$ will predominantly decay to
two photons, $a_X \rightarrow \gamma \gamma$, and could be easily distinguished
from the CP-even Higgs boson.

\section{The PQ symmetry with a singlet at the electroweak scale}

To set up a first version of the model we consider a set-up as given in
\cite{Kim84}. Later we shall simplify the model and restrict to the
fields that are relevant for the physics at the electroweak scale.
To break the PQ symmetry and SUSY, we generalize the Polonyi type
superpotential to break the PQ symmetry and parametrize the SUSY
breakdown. We introduce the following renormalizable
superpotential with the PQ symmetry and the U(1)$_R$ symmetry shown in
Table \ref{table:PQcharge},
\dis{
W=&   -H_uH_d X + mX \overline{X} -\eta \overline{X} S_1^2\\
&-\xi H_uH_d X' + m'X' \overline{X}\\
&+Z_1( S_1S_2  -F_1^2)     +Z_2(S_1S_2 -F_2^2)
}
where $F_1^2\ne F_2^2$ are constants. Here, we need $\langle  S_{1,2}\rangle=O(F_{1,2})$, but the rest is at the electroweak scale, \ie $ \langle  H_{u,d}\rangle=O(M_Z)$, $\langle  X,X'  \rangle  =O(M_Z)$, $\langle  Z_{1,2}\rangle=O(M_Z)$, and $\langle\overline{X} \rangle \le O(M_Z)$.

\begin{table}
\begin{center}
\begin{tabular}{c|cc|cc|cc|ccc}
\hline
 & $H_u$ & $H_d$ & $S_1$& $S_2$ & $Z_1$ & $Z_2$ &$X$ &$X'$ & $\overline{X}$ \\
\hline
$Q_{\rm PQ}$ & $+1$ & $+1$ & $-1$  & $+1$  & $0$ & $0$ &$-2$ &$-2$ &$+2$\\
\hline
$R$ & $+1$ & $+1$& $0$  & $0$  & $2$ & $2$ &$0$ &$0$ &$2$ \\
[0.3em]\hline
\end{tabular}
\caption{The PQ and $R$ charges of  $H_{u,d}, S_{1,2} Z_{1,2}, X$ and $\overline{X}$. }
\label{table:PQcharge}
\end{center}
\end{table}

The potential is
\dis{
V=&V_F+ V_D+ V_{\rm soft}.
}

The F-term potential is given by
\dis{V_F&=\Big| 
X+\xi X'  \Big|^2 (|H_u|^2+|H_d|^2)\\
&   +|-H_uH_d+m \overline{X}|^2+|-\xi H_uH_d +m' \overline{X}|^2\\
& +|\tilde{m} \tilde{X}-\eta S_1^2|^2 +|Z_1 +Z_2|^2|S_1|^2\\
&+\Big| 
-2\eta \overline{X}S_1+(Z_1+Z_2)S_2\Big|^2\\
&+|S_1S_2-F_1^2|^2+|S_1S_2-F_2^2|^2,\label{eq:VFmodel}
 }
where
\dis{
&\tilde{X}=\cos\alpha X+ \sin\alpha X',\\
& X_{\rm ew}=-\sin\alpha X+ \cos\alpha X',\\
& \cos\alpha=\frac{m}{\tilde{m}},~~ \sin\alpha=\frac{m'}{\tilde{m}},~~  \tilde{m}=\sqrt{m^2+m^{\prime\,2}}.
}
The D-term potential is given as usual,
and the TeV scale soft terms are
\dis{&V_{\rm soft}=
-m_{u}^2|H_u|^2+m_{d}^2|H_d|^2+M_1^2 |Z_1|^2+M_2^2 |Z_2|^2 \\
&~+m_1^2 |X|^2 +m_2^2 |X'|^2 +m_3^2 |\overline{X}|^2+\mu_1^2 |S_1|^2 +\mu_2^2 |S_2|^2.
}

The important terms determining the vacuum expectation values of $S_1, S_2$ and $\tilde{X}$ are
\dis{
V'=|S_1S_2-F_1^2|^2+|S_1S_2-F_2^2|^2 +|\tilde{m} \tilde{X}- \eta S_1^2|^2.\label{eq:s12xdetermine}
}
Let the phases of $S_1S_2$ and $\tilde{X}$ be $\delta_s$ and $\delta_{\tilde x}$, respectively. Then, $\delta_s=0$ and $\delta_s-2\delta_{\tilde x}=0$ determine
\dis{
s_1=\sqrt{\frac{mx}{\eta}}, ~~ \frac{s_2}{s_1}=\frac{\eta F^2}{2mx}
}
where $F^2=F_1^2+F_2^2, s_{1,2}=|S_{1,2}|$  and ${\tilde x}=|{\tilde X}|$. With $m=O(M_P)\sim O$(\MG) and ${\tilde x}=O(\tev)$, we obtain $\langle S_{1,2}\rangle$ at the intermediate scale. With $F_{1,2}$ at the intermediate scale, this scenario is realized.
Here, we note that $X_{\rm ew}$ does not appear in Eq. (\ref{eq:VFmodel}) and survives to the electroweak scale. Integrating out ${\tilde X}$, we consider the following terms in the superpotential
\dis{
W_{ew}= - \mu H_u H_d -f_h H_uH_d X_{\rm ew}
}
where
\dis{
f_h= -\sin\alpha +\xi\cos\alpha,
}
and soft terms of $H_u, H_d$ and $X_{\rm ew}$ (with a tadpole term possibility). The reason that one PQ charge carrying singlet survives below the axion scale comes from the fact that we have
one more field with charge $Q_{\rm PQ}=-2$  than fields with $Q_{\rm PQ}=+2$. This asymmetric appearance of the PQ fields is of general phenomena in string compactifications \cite{AxString}.

The same objective can be achieved with less fields but with the nonrenormalizable term,
\dis{
W=& -\frac{S_1^2}{M_P} H_uH_d - f_h H_uH_d X_{\rm ew}\\
&  +Z_1(S_1S_2 - F_1^2) + Z_2(S_1S_2 - F_2^2).\nonumber
}
While there are many ways to introduce the \NPQ\ at the electroweak scale,
one aspect is true for all of them: if the PQ symmetry forbids the $H_uH_d$ term
then the $\mu H_uH_d$ must appear by breaking the PQ symmetry at a high energy
scale. In addition, if a light singlet $X_{\rm ew}$ carrying the PQ charge $-2$
survives down to the electroweak scale, then the only additional
superpotential term is $X_{\rm ew} H_uH_d$, \ie the $X_{\rm ew}, X_{\rm ew}^2$ and $X_{\rm ew}^3$ terms
are not allowed with the minimal form of the K\"ahler potential.

If $X_{\rm ew}$ is charged under a new U(1)$'$ gauge symmetry which is broken at high energy scale, the D-term potential gives rise to a fine-tuning problem.Therefore, we do not introduce a quartic term from the D-term and stabilize $X_{\rm ew}$ by the positive quadratic term $m_e^2 |X_{\rm ew}|^2$.

If $f_h$ is large, the soft term of $X_{\rm ew}$ will be a subject of
renormalization group. For $f_h\sim 1$ and $A\sim 200$ GeV, we note that $m_e^2$ stays positive even with its vanishing GUT scale value, $m_{e\,\rm GUT}^2=0$.

\begin{widetext}
\section{Raising the Higgs mass}

With this effective superpotential and soft terms we can consider the following potential for Higgs and $X_{\rm ew}$:
\dis{V&=|\mu+f_h X_{\rm ew}|^2(|H_u|^2+|H_d|^2)+f_h^2|H_uH_d|^2
-m_u^2|H_u|^2+m_d^2|H_d|^2- (B\mu H_uH_d + {\rm c.c.})
\\
&+m_e^2|X_{\rm ew}|^2-(AX_{\rm ew}H_uH_d+ {\rm c.c.})
+\frac18(g_Y^2+g_2^2)(|H_u|^2-|H_d|^2)^2+\frac{g_2^2}{2}|H_u^\dagger H_d|^2
.}
Now, we can decompose neutral fields as real and complex components, $\phi=\frac{1}{\sqrt2}(\phi^r+i\phi^i)$ where $\phi=H_u^0, H_d^0, X_{\rm ew}$. At the vacuum, they take VEVs $v_u, v_d, x$, respectively, and
\dis{
V^{\rm min}&=\frac12[(\mu+\frac{f_h}{\sqrt2}x)^2-m_u^2]v_u^2+\frac12[(\mu+\frac{f_h}{\sqrt2}x)^2+m_d^2]v_d^2
\\
&+\frac{f_h^2}{4}v_u^2v_d^2+\frac{1}{32}(g_Y^2+g_2^2)(v_u^2-v_d^2)^2
 -B\mu v_uv_d-\frac{A}{\sqrt2}xv_uv_d +\frac12 m_e^2 x^2.\label{eqVmin}
}

Thus, the parameters we introduced for a fixed $X_{\rm ew}$ quantum number are $x$, $v$, $\tan\beta$, $f_h$, $\mu$, $A$, $B$,  $m_u^2$,  $m_d^2$ and  $m_{X_{\rm ew}}^2$.
Three minimization conditions of $V_{\rm min}$ of Eq. (\ref{eqVmin}) and $v\simeq$ 246 GeV reduce the number of independent parameters to six (seven).
The CP odd and even mass matrices are
\dis{
M^2_P=\left(
\begin{array}{ccc}
  (\frac{A}{\sqrt2}x+B\mu)\frac{v_d}{v_u}, & (\frac{A}{\sqrt2}x+B\mu), &\frac{A}{\sqrt2}v_d\\
  (\frac{A}{\sqrt2}x+B\mu), & (\frac{A}{\sqrt2}x+B\mu)\frac{v_u}{v_d}, & \frac{A}{\sqrt2}v_u \\
  \frac{A}{\sqrt2}v_d, & \frac{A}{\sqrt2}v_u, & M^2_O
  \end{array}\right)
  }
\dis{
M^2_H=
\left(
\begin{array}{ccc}
 \Big[\begin{array}{c}m_0^2 \cos^2\beta\\ +M_Z^2\sin^2\beta\end{array}\Big]
   & \Big[\begin{array}{c}\frac12 \sin 2\beta(f_h^2 v^2\\ -m_0^2-M_Z^2)\end{array}\Big] & \Big[\begin{array}{c}m_c^2 \sin \beta\\ -m_c^{ \prime 2} \cos \beta\end{array}\Big]\\
 \Big[\begin{array}{c}\frac12 \sin 2\beta(f_h^2 v^2\\ -m_0^2-M_Z^2)\end{array}\Big] &  \Big[\begin{array}{c} m_0^2 \sin^2\beta\\ +M_Z^2\cos^2\beta\end{array}\Big] & \Big[\begin{array}{c} m_c^2 \cos \beta\\
 -m_c^{ \prime 2} \sin \beta\end{array} \Big]\\
 \Big[\begin{array}{c}m_c^2 \sin \beta\\ -m_c^{ \prime 2} \cos \beta\end{array}\Big]
      & \Big[\begin{array}{c} m_c^2 \cos \beta\\
   -m_c^{ \prime 2} \sin \beta\end{array}\Big] & M_E^2
  \end{array}\right)\label{eq:HiggsMass}
  }
  \end{widetext}
where $M^2_O=\frac{1}{\sqrt2 x}(Av_uv_d-\mu f_h (v_u^2+v_d^2))$, $M^2_E= M^2_O$,
$m_c^2=f_h(\sqrt2 \mu +f_h x )v,~m_c^{\prime 2} =Av/\sqrt2$, and $m_0^2=  ( {\sqrt2}Ax+ 2B\mu)/\sin 2\beta$.

{\it Eigenvalues of $M_H^2$}: The smallest eigenvalue of CP even Higgs mass matrix $M_H^2$, Eq. (\ref{eq:HiggsMass}), is mostly $H_u-$like. Since (33) element  is inversely proportional to $x$, the smallest eigenvalue is close to the smallest eigenvalue of the $2\times 2$ submatrix composed of (11), (13), (31), and (33) elements for large $x$ whereas that of the $2\times 2$ submatrix composed of (11), (12), (21), and (22) elements,
\dis{
2 &m_h^{0\,2} \simeq (m_0^2+M_Z^2)- [ (m_0^2+M_Z^2)^2-4m_0^2 M_Z^2\\
& \cdot\cos^2 2\beta+f_h^2v^2(f_h^2v^2-2m_0^2-2M_Z^2)\sin^2 2\beta ]^{1/2}
}
for small $x$.

\begin{figure}[!t]
  \begin{center}
  \begin{tabular}{c}
   {\includegraphics[width=0.24\textwidth]{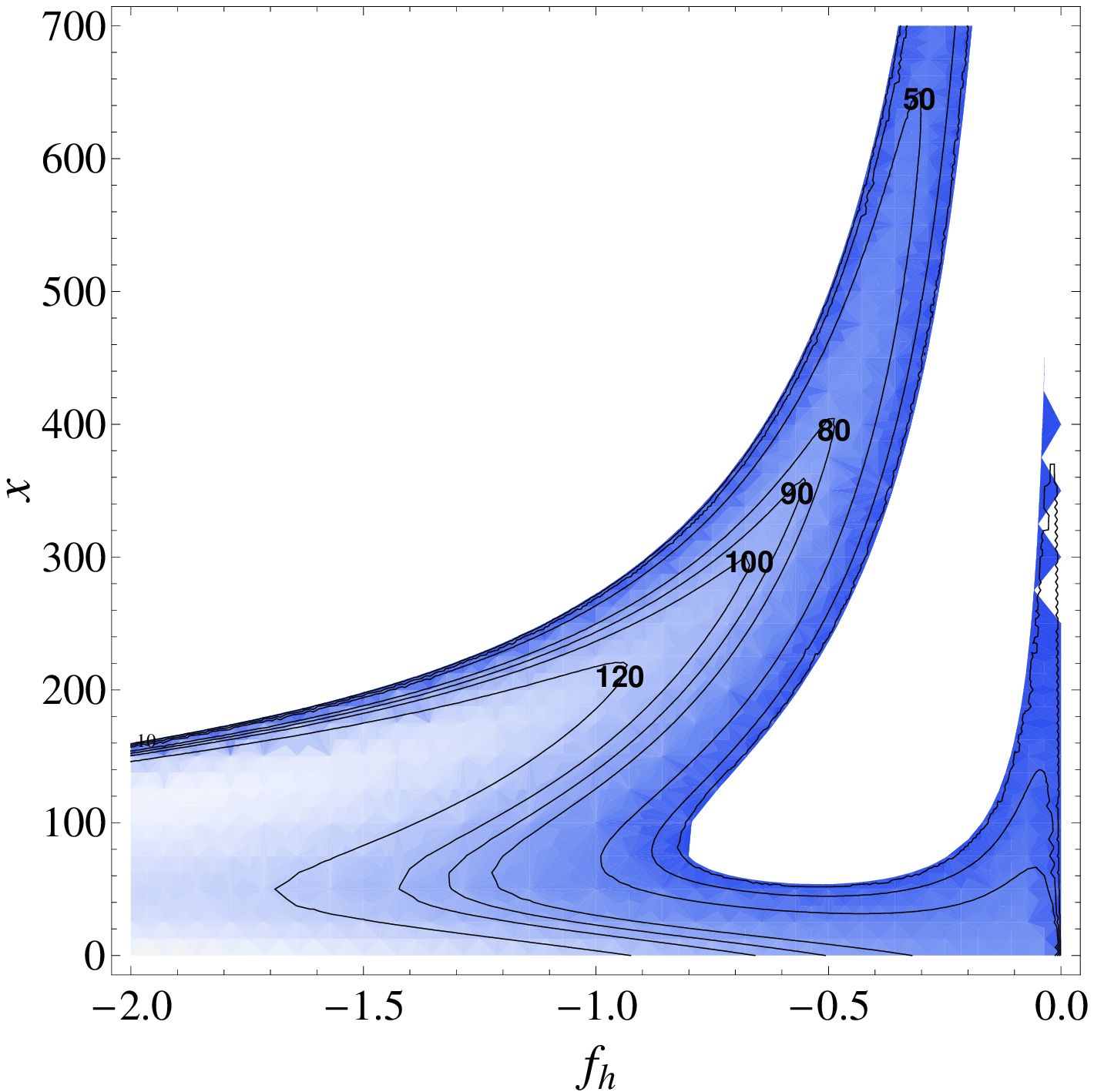}}
    {\includegraphics[width=0.24\textwidth]{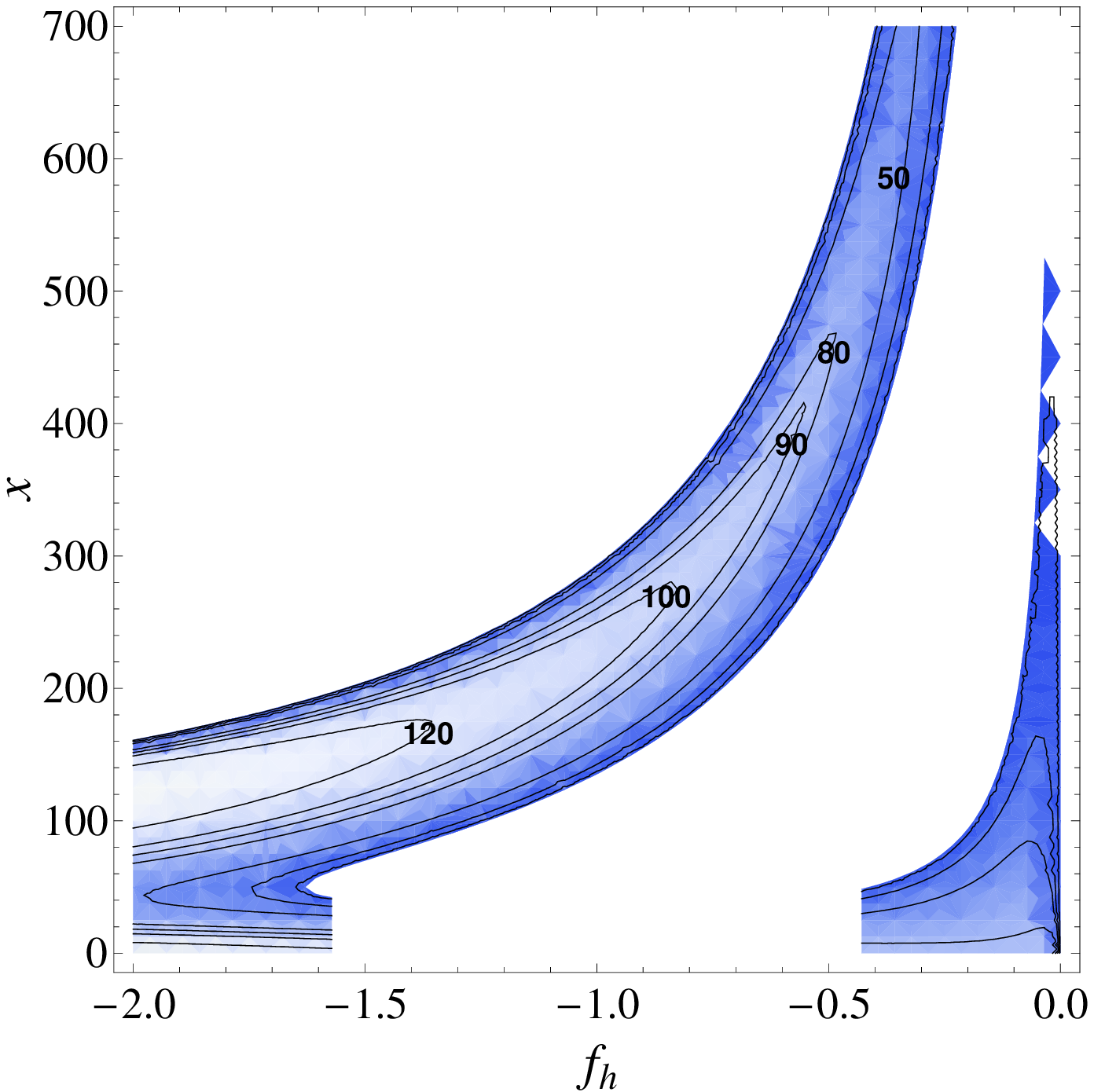}}
    \\
       {\includegraphics[width=0.24\textwidth]{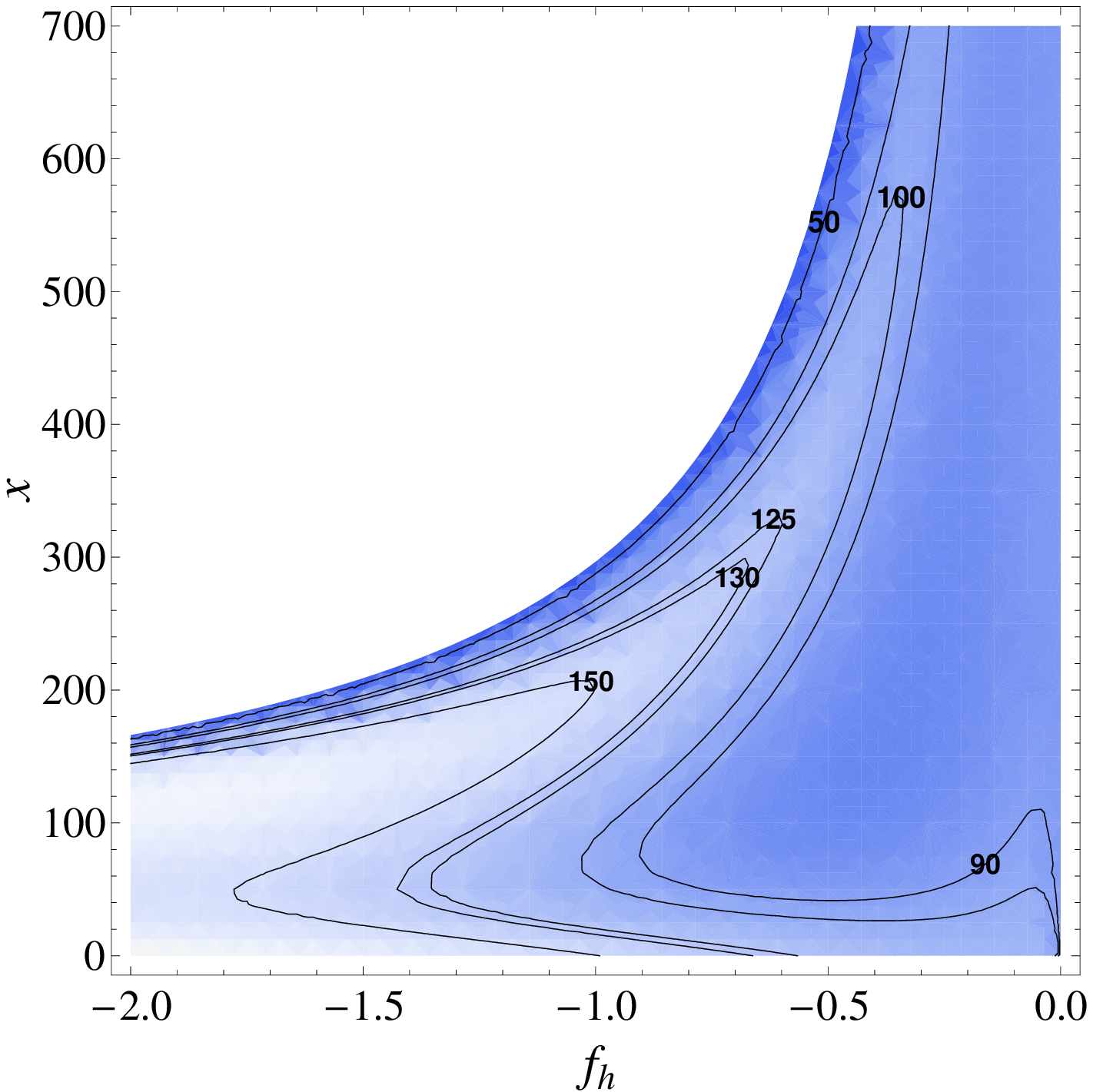}}
    {\includegraphics[width=0.24\textwidth]{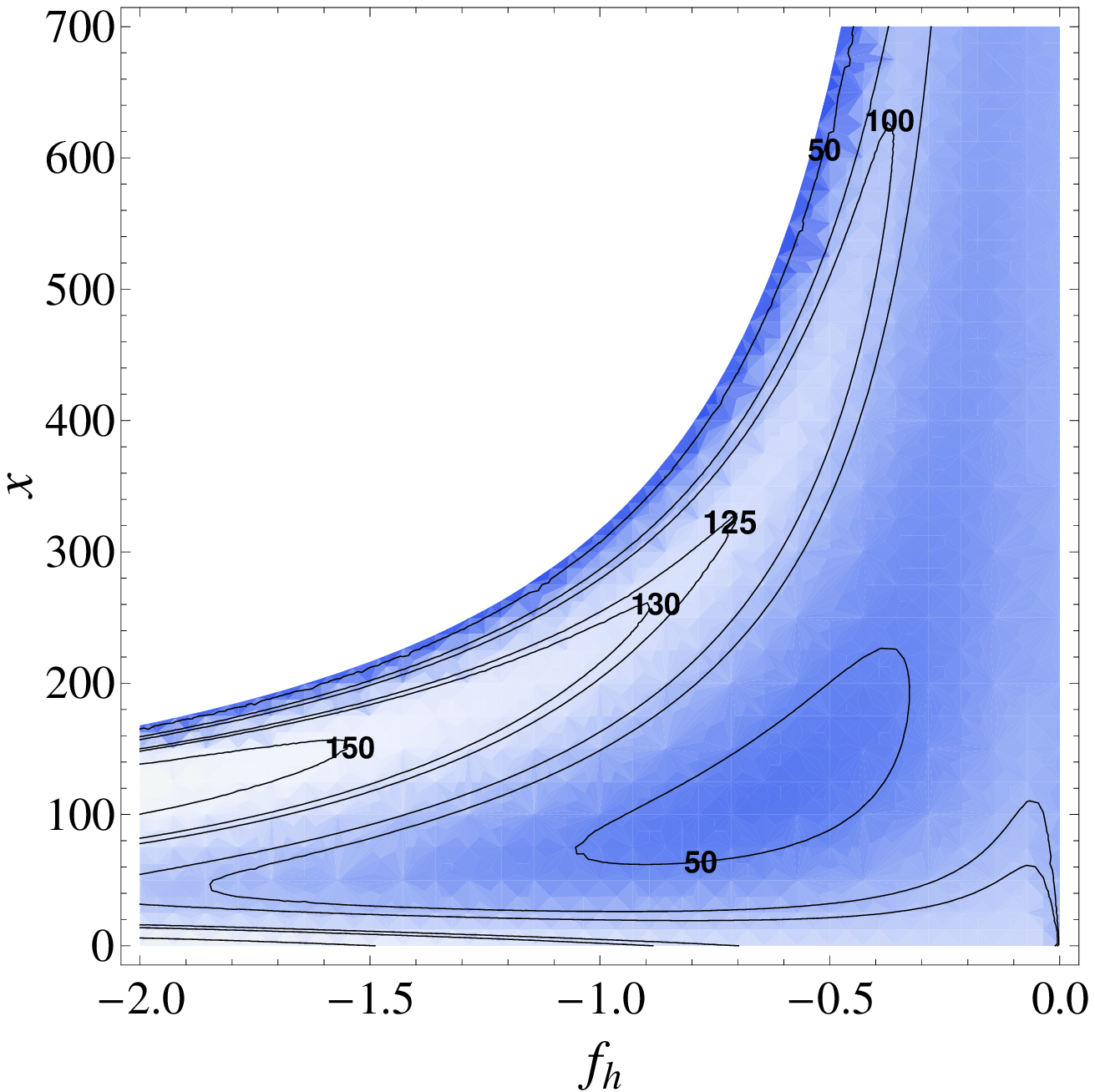}}
   \end{tabular}
  \end{center}
\caption{
The Higgs boson masses in the GeV units. The tree level masses are shown in (a) $\tan\beta=3$ and (b) $\tan\beta=5$. Masses with radiative corrections for $M_s=1\,\tev$, $A_t=800\,\gev$ are shown in (c) $\tan\beta=3$ and (d) $\tan\beta=5$. We use $B=500\,\gev$ and $A/f_h=\mu=200\,\gev$. } \label{fig:HiggsMass1}
\end{figure}

In Fig. \ref{fig:HiggsMass1}, we show the Higgs boson masses in the $x-f_h$ plane for $\tan\beta=3$[(a), (c)] and for $\tan\beta=5$[(b), (d)], with[(a), (b)] and without [(c),(d)] the radiative corrections.  We used $M_s=1\,$TeV, $A_t=800\,\gev$, and $B=500$GeV, $A/f_h=\mu=200~\gev$.
For the quantum corrections, we consider two more parameters: the geometric mean of the stop masses  $M_s=\sqrt{m_{\tilde t_1}m_{\tilde t_2}}$ and the $A_t$ term from the top Yukawa coupling. For $A_t=800\,\gev$ and $M_s=1\,\tev$, the radiative mass shift to the tree level mass $m_h^0\simeq 96\,\gev$  is about 30\,GeV.

If we require that perturbativity holds up to the PQ scale, we need $|f_h|\leq 0.8$.

\begin{figure}[!t]
  \begin{center}
  \begin{tabular}{c}
   \includegraphics[width=0.4\textwidth]{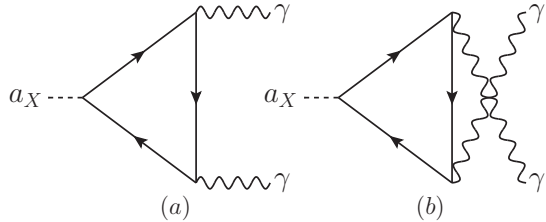}
   \end{tabular}
  \end{center}
 \caption{ The Feynman diagrams for the $a_X\gamma\gamma$ coupling with the Higgsino loop.
  }
\label{fig:Triangle}
\end{figure}

{\it Eigenvalues of $M_P^2$}: One eigenvalue of $M_P^2$ is 0, corresponding to the longitudinal component of $Z$ boson. Among the two remaining eigenvalues, the smaller one is
\dis{
2m^2_{a_X}= (m_0^2+M_O^2)- \Big[(m_0^2+M_O^2)^2
-\frac{4\mu \tilde{M}^3 }{\sin 2\beta}\Big]^{1/2}\label{eq:CPoddmass}
}
where $\tilde M^3 = 2BM_O^2-f_hA(v_u^2+v_d^2)$.
\begin{figure}[!t]
  \begin{center}
  \begin{tabular}{c}
   \includegraphics[width=0.24\textwidth]{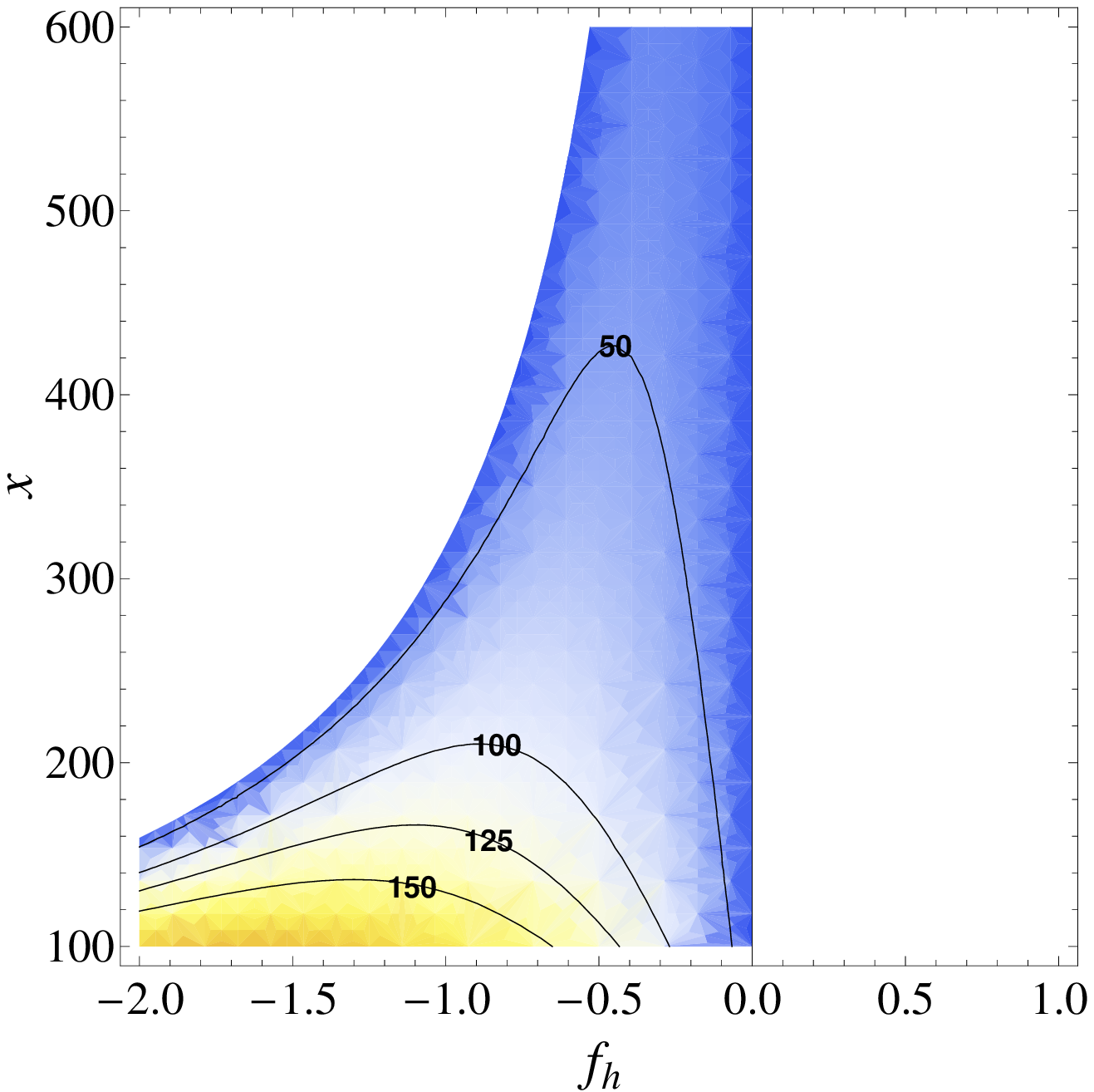}
   \includegraphics[width=0.24\textwidth]{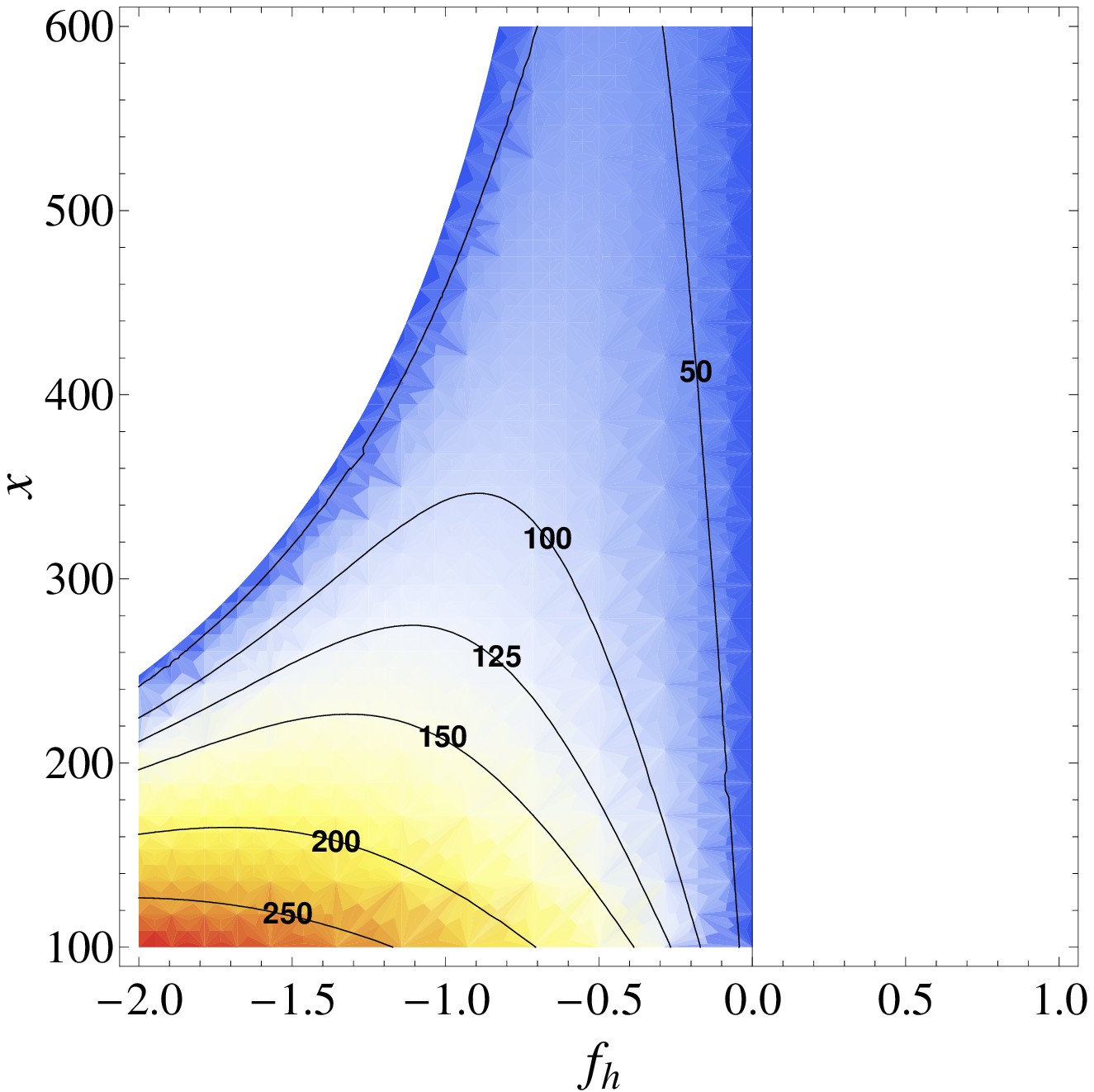}
   \end{tabular}
  \end{center}
 \caption{  The $a_X$ masses for $\tan\beta=3$: (a) $\mu=150\,\gev$ and (b) $\mu=200\,\gev$.}\label{fig:massCPodd}
\end{figure}
From Eq. (\ref{eq:CPoddmass}), we note that the $a_X$ mass is small for a small $\mu$. However, the $(1/\sin{2\beta})$ dependence is not singular for $\sin{2\beta}\to 0$ because the numerator cancels this divergence.
In Fig. \ref{fig:massCPodd}, the mass of $a_X$ is shown in the $x-f_h$ plane for $\mu=150\,\gev$ and 200\,GeV, respectively, and $A=f_h\times200\,\gev$, $B=500\,\gev$ and $\tan\beta=3$. In this parameter range, the lightest eigenvalue of $M_P^2$ is $X_{\rm ew}$-like.
\color{black}
As in the axion-photon-photon coupling case, in general there exists an $a_X$-photon-photon coupling as shown in Fig. \ref{fig:Triangle}. Since the diagram occurs through the Higgsino line only, the anomaly coupling estimation is simple to give
\dis{
{\cal L}_{a_X\gamma\gamma}=\frac{\alpha_{\rm em}}{4\pi} \frac{a_X}{x} F_{\rm em\,\mu\nu} {\tilde F}_{\rm em}^{\mu\nu} \label{eq:xAnomaly}
}
where $F_{\rm em\,\mu\nu}$ is the electromagnetic field strength and ${\tilde F}_{\rm em}^{\mu\nu}$ is its dual. We note that the coupling is not suppressed by the axion decay constant $F$ but by the VEV of $X_{\rm ew}$.\footnote{The term suppressed by $F$ is the axion-photon-photon coupling. The chiral current of the charged Higgsino, $J_5^\mu=\overline{\tilde H}\gamma^\mu\gamma_5 \tilde H$, has the divergence $\partial_\mu J_5^\mu=(\alpha_{\rm em}/2\pi) F_{\rm em\,\mu\nu} {\tilde F}_{\rm em}^{\mu\nu} +2\mu\overline{\tilde H}\gamma^\mu\gamma_5 \tilde H $ which gives Eq. (\ref{eq:xAnomaly}).
}
Similar anomaly couplings to $W_{\mu\nu} {\tilde W}^{\mu\nu} $ and $Z_{\mu\nu} {\tilde Z}^{\mu\nu} $ are present with couplings proportional to $\alpha_2$ and $\alpha_Z$, respectively. But, $a_X hh$ and $a_X H^+H^-$ are not present. Therefore, the production and decay of $a_X$ occur with the electroweak scale.
Through the anomalous coupling (\ref{eq:xAnomaly}), the electron-positron collider LEP II (with $\sqrt{s}\ge 130\gev$) could have produced three photon events (through $e^+e^-\to\gamma a_X$) for $125\,\gev$ pseudoscalar  at a $10^{-5}-10^{-6}$ pb level,

\dis{\sigma(s)&=\frac{3\alpha_{\rm em}^3}{256 \pi} \frac{1}{x^2}\Big(1-\frac{m_{a_X}^2}{2s}\Big)\Big(1-\frac{m_{a_X}^2}{3s}
+\frac{m_{a_X}^4}{12s^2}\Big),\label{eq:eepseudo}
}
which gives only $\sim 5\times 10^{-3}$ event for the integrated luminosity of LEP II \cite{LEPII}.

With Eq. (\ref{eq:xAnomaly}),  the decay width is given by
\dis{\Gamma(a_X \to \gamma \gamma)=\frac{\alpha_{\rm em}^2 m_{a_X}^3}{64 \pi^2 x^2 }.\label{eq:aXtotwoPh}
}
Since the LHC lower bound of the Higgsino mass is above 200 GeV \cite{LHCChargino},
the $a_X$ decay to two photons for its mass of order 125 GeV with the insertions of $f_h$ and $1/\mu$ is negligible compared to Eq. (\ref{eq:aXtotwoPh}).
For $m_{a_X}< 2M_W$, the decay $a_X\to W+{\rm lepton}+{\rm neutrino}$ introduces a suppression factor $m_{\rm lepton}^2q^2_W/M_W^4$. A similar remark applies to the $Z_{\mu\nu}\tilde{Z}^{\mu \nu}$ coupling.
On the other hand, some superpartner fermion can be lighter than $a_X$.

For the $X_{\rm ew}$-like $a_X$, therefore, decays to the $\gamma\gamma$ and a pair of lightest neutralino modes account for almost 100 \%.
If so, some two photon events with strong dijets in the forward direction, which is the characteristic of vector boson fusion, may come from $a_X$ production and decay, showing a two-photon resonance peak at the $a_X$ mass different from $125\,\gev$. Such event rate is too small to be observed at the LHC, since the ratio of the $a_X$ production to that of the MSSM Higgs boson $h$ is naively estimated by the ratio of couplings,  $\sim \alpha_2^4/\alpha_c^4\sim (0.0336/0.118)^4\simeq 0.65\times 10^{-2}$ \cite{couplings}.

Even though the lightest eigenvalue of $M_P^2$ is $X_{\rm ew}$-like, a small fraction of the Higgs component makes $a_X$ decay to $b\bar{b}$ at tree level.
The ratio of BRs to $a_X \to \gamma \gamma$ and to $a_x \to b\bar{b}$ in our example is given by
\dis{
R=\frac{\alpha_{\rm em}^2 m_{a_X}^3}{64 \pi^2 x^2 } \frac{8\pi M_W^2}{3C^2 m_b^2m_{a_X}}
=\frac{\alpha_{\rm em}^2 m_{a_X}^2}{ 24\pi x^2 }\frac{ M_W^2}{C^2 m_b^2},\label{eq:Ratiotob}
}
where $C=[g_2\tan\beta\tan\gamma- (M_W/x)]\cos\gamma $ with
$\tan\gamma=-(\cos\beta\cot\beta)/[(x/v)+ (\sqrt2B\mu/Av)]$ in a large $\tan \beta$ limit,
and $\frac13$ is multiplied for three colors of $b$. $R$ of Eq. (\ref{eq:Ratiotob})  is about $( m_{a_X}/67\,C x )^2$.
In a large $\tan \beta$ and a small $\mu$ limits, $C\approx -g_2v\cos\beta/x $ which is very small. Note in addition that the absolute production rate through Eq. (\ref{eq:aXtotwoPh}) is relevant at this stage as far as the $b\bar b$ mode is swamped by the SM background. If the production through the initial $q\bar q$ as in Eq. (\ref{eq:eepseudo}) is estimated for $10\, {\rm fb}^{-1}$ luminocity of the 7\,TeV LHC, we expect roughly $10^{-2}$ event.

\section{Conclusion}
Experiments at the LHC will soon test the properties of the SM Higgs boson \cite{LHC12}. Masses and decay properties of the Higgs system will be crucial for the analysis of potential physics beyond the standard model. In this work we have considered a specific scheme
motivated by supersymmetry and the strong CP-problem that predicts
a pseudoscalar particle with decay $a_X \to \gamma\gamma$, that might
well be within reach of current LHC experiments.

\vskip 0.2cm

\noindent {\bf Acknowledgments}: {We are grateful to  Bumseok Kyae, Hyun Min Lee, and especially to Stuart Raby for helpful discussions. JEK and MSS are supported in part by the National Research Foundation (NRF) grant funded by the Korean Government (MEST) (No. 2005-0093841), and HPN is partially supported by the SFB-Transregio TR33 ``The Dark Universe�� (Deutsche Forschungsgemeinschaft) and the European Union 7th network program ``Unification in the LHC era�� (PITN-GA-2009-237920).
}


\end{document}